\documentclass[submission,copyright,creativecommons]{eptcs}
\providecommand{\event}{FICS 2013} 
\usepackage{breakurl}             

\usepackage{amssymb,color,theorem,amsmath}
\usepackage{hyperref}

\usepackage{graphicx}

\newtheorem{definition}{Definition}[section]
\newtheorem{lemma}[definition]{Lemma}

\newtheorem{theorem}[definition]{Theorem}

\newcommand{\Atom}      { { \mathbb{A} } }
\newcommand{\State} {{\mathbb{S}} }

\newcommand{\True}	        {\mbox{\sf T}}

\newcommand{\Nat}           { { \mathbb{N} } }

\newcommand{\FinState}	{\State_{\mbox{\tiny fin}}}

\newcommand{\FinPow}		{{\cal P}_{\mbox{\tiny \it fin}}}

\newcommand{\Level}					{{\sf lev}}

\newcommand{\restr}					 {\upharpoonright}

\newcommand{\Truth}					{{\sf tr}}

\newcommand{\query}              { {\tt q}}

\newcommand{\Reduction}          { {\tt R} }

\newcommand{\reduces}            { {\twoheadrightarrow} }

\newcommand{\SN}               {{\tt SN}}
\newcommand{\Tree}               {{\tt T}}
\newcommand{\ON}               {{\tt ON}}
\newcommand{\next}               {{\tt next}}
\providecommand{\urlalt}[2]{\href{#1}{#2}} \providecommand{\doi}[1]{doi:\urlalt{http://dx.doi.org/#1}{#1}}

\title{Non-monotonic Pre-fixed Points and Learning}
\author{Stefano Berardi  \qquad Ugo de'Liguoro \\
\institute{Universit\`a di Torino}
\email{\quad stefano.berardi@unito.it \qquad ugo.deliguoro@unito.it}}

\def\titlerunning{Non-monotonic Pre-fixed Points and Learning}
\def\authorrunning{S. Berardi \& U. de' Liguoro}

\begin{document}
\maketitle

\begin{abstract}
We consider the problem of finding pre-fixed points of interactive realizers over arbitrary knowledge spaces, obtaining a relative recursive procedure.
Knowledge spaces and interactive realizers are an abstract setting to represent learning processes, that can interpret non-constructive proofs. Atomic pieces of information of a knowledge space are stratified into levels, and evaluated into truth values depending on knowledge states. Realizers are then used to define operators that extend a given state by adding and possibly removing atoms: in a learning process states of knowledge change non-monotonically.
Existence of a pre-fixed point of a realizer is equivalent to the termination of the learning process with some state of knowledge which is free of patent contradictions and such that there is nothing to add.
In this paper we generalize our previous results in the case of level 2 knowledge spaces and deterministic operators to the case of $\omega$-level knowledge spaces and of non-deterministic operators.
\end{abstract}

\section{Introduction}
\label{section-introduction}

A fundamental aspect of constructive interpretations of classical arithmetic is how information is gathered and handled while looking for a witness of the proved formulas. This has been understood by several authors as a problem of control and side effects, although intended in different ways. Building over Coquand's semantics of evidence of classical arithmetic \cite{Coq95}
and its representation as limiting interaction sequences \cite{BerardiLiguoro},
we have developed the concept of {\em interactive realizability} in
\cite{AB,BerardiLiguoroMonadi}, which consists of interpreting non constructive proofs as effective strategies that ``learn''  the  witness.

According to \cite{full-abstract}, learning the truth of an arithmetical statement can be abstractly presented as a process going through steps, which we call {\em states of knowledge}, such that a (candidate) witness can be relatively computed out of them. These are certain subsets of a countable set $\Atom$ whose elements are pieces of evidence that we dub {\em answers}.
On the other hand $\Atom$ is equipped with an equivalence relation $\sim$ whose equivalence classes $[a]_{\sim}$ are {\em questions}; since we allow that in a state of knowledge each question has at most one answer, we say that $X$ is a state if
for all $a\in\Atom$, the set $X \cap [a]_{\sim}$ is either a singleton or empty. We also denote by $\State$ the set of states.

Over states we can define a ``query map'' $\query([a]_{\sim},X) \in \FinPow(\Atom)$, taking a question $[a]_{\sim}$, a state $X \in \State$, and returning the set $X\cap[a]_{\sim}$, that is a singleton $\{b\}$ if $b \in X$ is  the only answer to $[a]_{\sim}$; the empty set otherwise. We call {\em state topology} the smallest topology making the query map continuous. Equivalently the state topology is generated by the canonical sub-basics $A_a = \{X \in \State \mid a \in X\}$ and $B_a = \{X \in \State \mid X \cap [a]_{\sim} = \emptyset \}$ for $a\in\Atom$.

Knowledge is improved by means of ``realizers'' $r : \State \rightarrow \FinPow(\Atom)$ that are functions guessing a finite set of new information $r(X)$ with respect to the current state of knowledge $X$. We assume that $r(X)\subseteq\Atom$ is always a finite set, so that a step of an ``algorithm'' to compute with $r$ consists of proceeding from some $X$ to $X' \cup Y$, that we treat here as a reduction relation $X \reduces^r_1 X' \cup Y$, where $X'\subseteq X$ and $\emptyset \neq Y\subseteq r(X) \setminus X$ have to satisfy certain requirements. Under this respect if $r(X) \subseteq X$, namely $X$ is a pre-fixed point of $r$, then  the computation terminates in the state $X$.

In \cite{AB,BerardiLiguoroMonadi} we have studied the case where $\Atom$ is essentially made of decidable arithmetical statements which are known to be true, and considered the case where $r(X)$ is either a singleton or it is empty. In this case
$X \reduces^r_1 X \cup r(X)$ if $r(X)\neq\emptyset$, and the sequence of reductions $X_0 \reduces^r_1 X_1 \reduces^r_1 \cdots$ out of some $X_0$ is uniquely determined by $r$ and the sequence $X_0 \subseteq X_1 \subseteq \cdots$ is monotonic. Hence we have proved termination by applying Knaster-Tarski theorem.

We call deterministic the case in which $r(X)$ is at most a singleton. A first generalization of the picture is when $r(X)$ may include more than one answer, which is the non-deterministic case. Then $r(X)$ is not required to be a state, and the next state is $X \cup Y$, for some non-deterministic choice of a subset $Y\subseteq r(X)$ of pairwise unrelated answers w.r.t. $\sim$. A further extension  is when $X'$ is a proper subset of $X$ in the reduction step $X \reduces^r_1 X' \cup Y$, then loosing the monotonicity of the sequence $X_0, X_1, \ldots$. This is the case when the truth values of answers are logically related, and adding some new answer may turn to false the truth values of some previously true answers. In this case whenever we add some answer we have also to remove some, and the fixed point result becomes difficult to prove.


To model logical dependencies of answers we assume that $\Atom$ is ``stratified'' by a map $\Level : \Atom \rightarrow \Nat$, splitting the answers into $\omega$ levels, in decreasing order of ``reliability''. As we explained in \cite{AB,BerardiLiguoroMonadi}, we need $\omega$-levels of answers to describe the constructive content of classical proofs of arithmetic. Logical dependence means that an answer of level $n$ (e.g. a universal statement) that has been considered as true so far, might be falsified by discovering that an answer of level $< n$ (a counterexample) should be true. Hence we relativize the truth value of answers to a state (to which they do not necessarily belong) using a function $\Truth(a,X)$ that only depends on the answers in $X$ having a smaller level, that is $\Truth(a,X) = \Truth(a,\{x \in X \mid \Level(x) < \Level(a)\})$. Further we require that $\Truth(a,X)$ depends continuously on the state parameter w.r.t. the state topology. This is how we abstractly capture the idea that this should be a relative computable function, which will be recursive in case of a finite set $X$ of answers. Instead, we add {\em no} level restriction on a realizer $r$: if $X \in \State$, then the answers of level $n$ in $r(X)$ may depend on the answers of {\em any} level in $X$, including the answers of level $\ge n$ in $X$.
Finally we also say that $X \in \State$ is {\em sound} if $\Truth(a,X) = \True$ for all $a\in X$. Only sound pre-fixed points are of interest.

The fact that the truth value of an answer w.r.t. a state $X$ only depends on truth values of lower level answers in $X$
suggests the following non-deterministic algorithm to find a sound pre-fixed point of the function $r$: we pick one or more answers with the same level $n$ from $r(X)$ and dropping all answers of level $>n$ from $X$. We express the algorithm through the relation $X \reduces^r_1 X' \cup Y$ whenever $X' = \{x \in X \mid \Level(x) \le n \}$ and $Y \subseteq r(X)$ is a finite {\em homogeneous} state made of answers of the same level, say $n$, which is considered as the level of the state. Then we establish the main result of the paper, namely that if $r$ is a {\em realizer} (see Definition \ref{def:realizer} below) then any reduction $\reduces^r$ out of some sound $X_0$ terminates, within a finite number of steps, by a sound pre-fixed point of $r$, which is finite if $X_0$ is such.

We have a final warning about the proof in this paper. It is possible to show that our termination result implies the $1$-consistency of First Order Arithmetic, and therefore it is not provable in it. Thus, no elementary proof of our result is possible, although we have found several non-elementary proofs. The proof included here is classical and it uses set theory, choice axiom and uncountable reduction sequences: none of them is strictly required, but we trade off logical complexity for readability. We could remove ordinals, choice axiom and even Excluded Middle from the proof, at the price of a harder (and longer) argument.

\medskip
The plan of the paper is as follows. In \S \ref{section-algorithm} we define a reduction relation on states depending on a realizer $r$, which is the non-deterministic algorithm to search a pre-fixed point of $r$. In \S \ref{section-open} we prove that the set of states from which this algorithm always terminates is an open set in the state topology. In \S \ref{section-omega1} we use this fact to prove that if there is some  reduction sequence of length $\omega$ out of some state, then there is a reduction sequence of  length $\omega_1$ out of the same state. Eventually, in \S \ref{section-termination}, we prove that reduction sequences of length $\omega_1$  do not exist, so that we conclude that all reduction sequences of our algorithm are of finite length. Then in \S \ref{relatedWorks} we discuss some related works and we conclude.

\section{A non-deterministic parallel algorithm for finding pre-fixed points}
\label{section-algorithm}

For convenience we recall the basic definitions from \cite{full-abstract} and the introduction above. We are given a countable set $\Atom$ and an equivalence relation $\sim$ over $\Atom$; the map $\Level:\Atom \rightarrow \Nat$ respects $\sim$ that is $\Level(x) = \Level(y)$ if $x \sim y$; $X\subseteq \Atom$ is a state if for all $x,y \in X$, $x \neq y$ implies $x \not\sim y$; the set $\State$ of states is taken with the state topology, generated by the sub-basics $A_a = \{X \in \State \mid a \in X\}$ and $B_a = \{X \in \State \mid X \cap [a]_{\sim} = \emptyset \}$; we take $\Atom$ and $2$ with the discrete topology and $\Atom\times\State$ with the product topology.

\begin{definition}[Layered Valuation, Sound State and Realizer]\label{def:realizer}
A {\em layered valuation} over $(\Atom, \sim, \Level)$, shortly a {\em valuation}, is a continuous mapping $\Truth:\Atom \times \State \rightarrow 2$ such that
\[\Truth(a,X) = \Truth(a,\{x \in X \mid \Level(x) < \Level(a)\}).\]
A state $X\in \State$ is {\em sound} if $\Truth(x,X) = \True$ for all $x\in X$.

A {\em realizer} w.r.t. the valuation $\Truth$ is a continuous map $r:\State \rightarrow \FinPow(\Atom)$, where $\FinPow(\Atom)$ is taken with the discrete topology, which is such that:
\[\forall X \in \State\; \forall a \in r(X).~ X \cap [a]_{\sim} = \emptyset \And \Truth(a,X) = \True.\]
\end{definition}

 Given $n\in \Nat$ and a state $X$ we define the subsets of $X$:
\[X \restr_{< n} \;=\; \{x \in X\mid\Level(x) < n\}, ~~~ X \restr_{>n} \;=\; \{x \in X \mid \Level(x) > n\}, ~~~ X \restr_{= n} \;=\; \{x \in X \mid \Level(x) = n\}.\]
We also write $X \restr_{\le n} \;=\; X \restr_{< n} \;\cup\; X \restr_{= n} $.
We denote by $\FinState$ the set of finite states; let $s, s', t, t', \ldots$ range over $\FinState$.

\begin{definition}[Reduction]\label{def:reduction}
We say that a state $s\in\FinState$ is {\em homogeneous} if $s\neq\emptyset$ and for some $n\in\Nat$, $\Level(x) = n$ for all $x\in s$; then we write
$\Level(s) = n$. For any homogeneous $s$ of level $n$ we define a map $\Reduction_s: \State \rightarrow \State$ by:
\[\Reduction_s(X) \restr_{<n} \;=\; X \restr_{<n},~~~\Reduction_s(X) \restr_{=n} \;=\; X \restr_{=n} \cup\; s, ~~~ \Reduction_s(X) \restr_{>n} \;=\; \emptyset.\]
Then, given a realizer $r$ and an homogeneous $s$ we define the binary {\em reduction relation} over $\State$ by:
\[X \reduces^{s,r} Y \Leftrightarrow s \subseteq r(X) \And \Reduction_s(X) = Y.\]
\end{definition}


We say that {\em $X$ reduces to $Y$ in one step} and we write $X \reduces^{r}_1 Y$ if $X \reduces^{s,r} Y$ for some homogeneous $s$. As immediate consequence of the definitions of $\reduces^{r}_1$,  $\Truth$ and $r$ we establish:

\begin{lemma}\label{lem:pre-fix}
\hfill
\begin{enumerate}
\item $X \reduces^{r}_1 Y \And X \in \FinState \Rightarrow Y \in \FinState$.
\item $X \reduces^{r}_1 Y \And \;\mbox{$X$ is sound} \Rightarrow \mbox{$Y$ is sound}$.
\item $\neg \exists\, Y. ~ X \reduces^{r}_1 Y \Leftrightarrow r(X) \subseteq X$.
\end{enumerate}
\end{lemma}


A reduction sequence of length $n$ from $X$ to $Y$ is a sequence $X_0, \ldots, X_n$ such that $X = X_0 \reduces^{r}_1 X_1 \reduces^{r}_1 \ldots$ $ \reduces^{r}_1 X_n = Y$. An infinite reduction sequence out of $X$ is an endless sequence $X = X_0 \reduces^{r}_1 X_1 \reduces^{r}_1 \ldots \reduces^{r}_1 X_n \ldots $ of reductions. For any integer $n \in \Nat$ we say that $X$ reduces to $Y$ in $n$ steps and we write $X \reduces^{r}_n Y$ if there is a length $n$ reduction sequence from $X$ to $Y$. We write $X \reduces^{r} Y$ if $X \reduces^{r}_n Y$ for some $n \in \Nat$.

We observe that $X$ is a pre-fixed point of $r$, that is $r(X) \subseteq X$, if and only if there is no homogeneous set $s \subseteq r(X)$ such that $X \cap s = \emptyset$, that is if and only if for all $Y \in \State$ we have $X \not\! \reduces^{r}_1 Y$. If $\sim$ is decidable and both $r$ and $\Truth$ are relative recursive then we can see $Y \reduces^{r}_1 Z$ as the one step relation of a non-deterministic algorithm computing
a pre-fixed point $X$ of $r$ starting with some $X_0\in \State$; then such an $X$, if any, can be seen as a result of the computation starting with $X_0$. By lemma \ref{lem:pre-fix} we know that if we move from some finite sound state $s_0$, e.g. $\emptyset$, the reduction relation $\reduces^{r}_1$ generates a tree with finite and sound states as nodes, which is finitary because $r(X)$ is finite even for infinite $X$ so that there can be only finitely many homogeneous $s\subseteq r(X)$. In particular the relation  $X \reduces^{r}_1 Y$ is decidable for finite $X$ and $Y$, and relative recursive in general.

We say that $X \in \State$ is strongly normalizing if all reduction sequences out of $X$ are finite. We denote by $\SN$ the set of all strongly normalizing states. Our thesis is that $\SN = \State$, namely that the reduction tree out of any $X$ is finite. This implies that if $s\in\FinState$ and $s$ is sound we can effectively find a finite and sound pre-fixed point $t$ of $r$ by reducing $s$.

\section{The set of strongly normalizing states is open}
\label{section-open}
The first step toward establishing $\SN = \State$  is to prove that $\SN$ is open in the state topology. To prove this we first characterize the reduction relation.

\begin{lemma}[Reduction] \label{lemma-reduction}
Let $s \in \FinState$ be any homogeneous state of level $n$. Assume $X, Y \in \State$ and $X \reduces^{s,r} Y$. Let $m \in \Nat$.
\begin{enumerate}

\item 
$X \restr_{= n} \subset Y \restr_{= n}$

\item 
$X \not \reduces^{s,r} X$.

\item 
If $m \le n$, then $X \restr_{< m+1} \subseteq Y \restr_{<m+1}$

\item 
If $X \restr_{< m+1} \not \subseteq Y \restr_{<m+1}$, then $Y \restr_{= m} = \emptyset$.

\item 
If $X \restr_{< m} = Y \restr_{< m}$ then $m \le n$.

\item 
If $X \restr_{< m} = Y \restr_{< m}$ then $X \restr_{< m+1} \subseteq Y \restr_{<m+1}$
\end{enumerate}
\end{lemma}

{\bf Proof}
\begin{enumerate}
\item 
By definition of $X \reduces^{s,r} Y$ we have $s \not = \emptyset$, $X \cap s = \emptyset$ and $Y \restr_{= n} = X \restr_{= n} \cup s$. We conclude $X \restr_{= n} \subset Y \restr_{= n}$.
\item 
By point $1$, if $X \reduces^{s,r} Y$ then $X \restr_{= n} \subset Y \restr_{= n}$, hence $Y \not = X$.
\item 
Assume $m \le n$ in order to prove $X \restr_{< m+1} \subseteq Y \restr_{<m+1}$. We reason by cases.
\begin{enumerate}
\item
Let $m < n$. Then $m+1 \le n$. By definition of $X \reduces^{s,r} Y$ we have $X \restr_{< n} = Y \restr_{<n}$, and from $m+1 \le n$ we conclude $X \restr_{<m+1} = (X \restr_{< n})\restr_{<m+1} = (Y \restr_{<n}) \restr_{<m+1} = Y \restr_{<m+1}$.
\item
Let $m = n$. Then by point $1$ above and $X \restr_{< n} = Y \restr_{<n}$ we have $X \restr_{<m+1} = X \restr_{<n+1} \subset Y \restr_{<n+1} = Y \restr_{<m+1}$
\end{enumerate}

\item 
By point $3$, if $X \restr_{< m+1} \not \subseteq Y \restr_{<m+1}$, then $m > n$. We deduce $Y \restr_{= m }
\subseteq Y \restr_{>n} = \emptyset$.

\item 
Assume $X \restr_{< m} = Y \restr_{< m}$ in order to prove that $m \le n$. If it were $m > n$, we would deduce $ X \restr_{= n} = (X \restr_{< m}) \restr_{= n} = (Y \restr_{< m}) \restr_{= n} = Y \restr_{= n}$, contradicting point $1$. Thus, $m \le n$.

\item 
We apply points $5$ and $3$ in this order.
\end{enumerate}

The next step is to prove that $\SN$ is open in the state topology. For all $n \in \Nat$, $n>0$ we define $\SN_n = \{X \in \State |\forall Y\in\State. X\not \reduces^{r}_n Y\}$ the set of states from which there is no reduction sequences of length $n$ from $X$. The reduction tree $\Tree(X) = \{Y \in \State | X \reduces^{r} Y \}$ from $X \in \State$ is finitely branching: from any node $Y$, the number of children of $Y$ has upper bound the number of subsets of $r(Y)$, which is finite. By K\"{o}nig's Lemma, $\Tree(X)$ is finite if and only if all branches of tree (all reduction sequences from $X$) are finite. Thus, $\Tree(X)$ is finite if and only if there is some upper bound $n \in \Nat$ to the reduction sequences from $X$. This implies $\SN = \bigcup_{n \in \Nat} \SN_n$. Therefore in order to prove that $\SN$ is open it is enough to prove that all $\SN_n$ are open.


\begin{lemma}[$\SN$ is open]
\label{lemma-open}
Assume $s \in \FinState$ is any homogeneous state. Let $I, I_0, I_1 \in \FinPow(\Atom)$ be finite sets of answers. Assume $X, Y, X', Y' \in \State$.
\begin{enumerate}
\item
For all $a \in \Atom$, $\{X \in \State | a \not \in X \}$ is open.

\item
If $(I_0, I_1)$ is a partition of $I$, then $\{X \in \State | (I \cap X = I_0) \wedge (I \setminus X = I_1) \}$ is open.

\item
$\Reduction_s:\State \rightarrow \State$ is a continuous map.


\item
$\SN_1$ is open.

\item
For all $n \in \Nat$, $\SN_n$ is open.

\item
$\SN$ is open
\end{enumerate}
\end{lemma}

{\bf Proof}
\begin{enumerate}
\item 
Assume $a \in \Atom$ and $O = \{X \in \State | a \not \in X \}$. The set $O$ consists of all states including some element of $[a]_{\sim}$ different from $a$, or having empty intersection with $[a]_{\sim}$. Thus $O$ is the union of all sets $\{X \in \State | b \not \in X \}$ for $b \in [a]_{\sim}$ and $b \not a$, and of the set $\{X \in \State | X \cap [a]_{\sim} = \emptyset \}$. All these sets are basic open of the state topology, therefore $O$ is an open set of the state topology.

\item 
Assume that $(I_0, I_1)$ is a partition of $I$ and $O = \{X \in \State | (I \cap X = I_0) \wedge (I \setminus X = I_1) \}$. Since both $(I \cap X, I \setminus X)$ and $(I_0, I_1)$ are partitions of $I$, the condition $(I \cap X = I_0) \wedge (I \setminus X = I_1)$ is equivalent to $(I \cap X \supseteq I_0) \wedge (I \setminus X \supseteq I_1)$. Thus, $O$ is equal to the intersection of all sets $\{X \in \State | a \in X\}$, for any $a \in I_0$, and of all sets $\{X \in \State | a \not \in X\}$, for $a \in I_1$. These sets are finitely many because $I$ is finite, and are either sub-basic open, or are open by point $1$ above. Thus, $O$, being a finite intersection of open sets, is open.

\item 
Assume $s \in \FinState$ is an homogeneous state of level $n$. Assume $a \in \Atom$ and $A_a \{Z \in \State | a \in Z\}$, $B_a = \{Z \in \State | Z \cap [a]_{\sim}\}$ are sub-basic open. We have to prove that if $Y \in A_a$ then $\Reduction_s^{-1}(A_a)$, $\Reduction_s^{-1}(B_a)$ are open sets. We prove this statement by case analysis.
\begin{enumerate}
\item
If $\Level(a) > n$ then $\Reduction_s^{-1}(A_a) = \emptyset$.
\item
If $a \in s$ then $\Reduction_s^{-1}(A_a) = \State$.
\item
If $\Level(a) \le n$ and $a \not \in s$ then $\Reduction_s^{-1}(A_a) = A_a$.
\item
If $\Level(a) > n$ then $\Reduction_s^{-1}(B_a) = \State$.
\item
If $s \cap [a]_{\sim} \not = \emptyset$ then $\Reduction_s^{-1}(B_a) = \emptyset$.

\item
If $\Level(a) \le n$ and $s \cap [a]_{\sim} = \emptyset$ then $\Reduction_s^{-1}(B_a) = B_a$.
\end{enumerate}

\item 
$\SN_1$ is the set of states reducing to no state, equivalently, the set of states $X \in \State$ which are pre-fixed points of $r$. Thus, we have to prove that if $X$ is a pre-fixed point of $r$, then there is some open set $X \in O$ such that all $Y \in O$ are pre-fixed points of $r$. Let $O' = r^{-1}(\{r(X)\})$, $O'' = \{Y \in \State| (r(X) \cap Y = r(X)) \wedge (r(X) \setminus Y = \emptyset) \}$, and $O = O' \cap O''$. $O'$ is open because $r : \State \rightarrow \FinPow(\Atom)$ is continuous and $ \FinPow(\Atom)$ has the discrete topology. $O''$ is open by point $2$, with $I_0 = r(X)$ and $I_1 = \emptyset$. Thus, $O$ is open. By definition, $X \in O' = r^{-1}(\{r(X)\})$ and $X \in O'' = \{Y \in \State| Y \cap r(X) = r(X) \wedge Y \setminus r(X) = \emptyset\}$, because $r(X) \subseteq X$. Thus, $X \in O$. For any $Y \in O$ we have by definition of $O$: $r(Y) = r(X)$ and $r(Y) = r(X) \subseteq Y$, as we wished to show.

\item 
We prove that $\SN_n$ is open by induction over $n \in \Nat, n>0$. The case $n=1$ is the previous point. Assume $\SN_n$ is open in order to prove that $\SN_{n+1}$ is open. Let $X \in \SN_{n+1}$: we have to prove that there is some open set $X \in O \subseteq \SN_{n+1}$. $r(X) \setminus X$ is finite, therefore there are finitely many homogeneous states $s_1, \ldots, s_k \subseteq r(X) \setminus X$. These states define exactly all reductions from $X$: $X \reduces^{s_i,r}_1 X_i$, for $i = 1, \ldots, k$. From $X \in \SN_{n+1}$ we deduce $X_i \in \SN_n$ for all $i = 1, \ldots, k$. Let $O_i = \Reduction_{s_i}^{-1}(\SN_n)$: $O_i$ is open by point $3$ above, and $X \in O_i$ because $\Reduction_{s_i}(X) \in \SN_{n}$ by the assumption $X \in \SN_{n+1}$. Let $O' = r^{-1}(\{r(X)\})$, $O'' = \{Y \in \State | (r(X) \cap Y = r(X) \cap X) \wedge (r(X) \setminus Y = r(X) \setminus  X) \}$. By definition we have $X \in O'$, $X \in O''$. $O'$ is open because $r$ is continuous, and $O''$ is open by point $2$. Let $O = O' \cap O'' \cap O_1 \cap \ldots \cap O_n$: then $X \in O$ and $O$ is open. For all $Y \in O$ we have $r(Y) = r(X)$, and $r(Y) \setminus Y = r(X) \setminus Y = r(X) \setminus X$. Therefore the reductions from $Y$ are exactly in number of $k$: $Y \reduces^{s_i,r}_1 Y_i$ for all $i = 1, \ldots, k$. We have $Y_i \in \SN_n$ by $O \subseteq O_i = \Reduction_{s_i}^{-1}(\SN_n)$. We conclude that $Y \in \SN_{n+1}$, as wished.

\item 
$\SN$ is the union of all $\SN_n$, therefore is a union of open sets and it is open.

\end{enumerate}

\section{Reduction sequences of transfinite length}
\label{section-omega1}
The next step is to prove that if there are states in $\State \setminus \SN$, then there are reduction sequences of any transfinite length. From this fact we will derive a contradiction.

We denote the class of ordinals with $\ON$, and ordinals with Greek letters $\alpha, \beta, \gamma, \lambda, \mu, \ldots$. We recall that a limit ordinal is any ordinal $\lambda$ such that for all $\alpha < \lambda$ we have $\alpha + 1 < \lambda$. $\omega$, the first infinite ordinal, and $\omega_1$, the first uncountable ordinal, are limit. $\omega_1$ has the additional property that any l.u.b. of some countable set $I$ of ordinals all $< \omega_1$ is some $\xi < \omega_1$.

A sequence of length $\alpha$ on $\State$ is any map $\sigma : [0, \alpha[ \rightarrow \State$. We represent sequences of length $\alpha$ with indexed sets $\sigma = \{X_{\beta} | \beta < \alpha\}$. When $\alpha =$ some limit ordinal $\lambda$, the limit of a sequence $\{X_{\beta} | \beta < \lambda\}$ is defined as $\lim_{\beta \rightarrow \lambda} X_\beta = \cup_{\beta < \lambda} \cap_{\beta \le \gamma < \lambda} X_\gamma$. To put otherwise, $\lim_{\beta \rightarrow \lambda} X_\beta$ consists of all answers which belong to the states of $\{X_{\beta} | \beta < \lambda\}$ from some $\beta$ on. A {\em limit sequence of length $\alpha$} is any sequence $\{ X_{\beta} | \beta < \alpha \}$ of length $\alpha$ such that for all limit ordinal $\lambda < \alpha$ we have $X_{\lambda} = \lim_{\beta \rightarrow \lambda} X_{\beta}$. A {\em limit reduction sequence of length $\alpha$} is any limit sequence of length $\alpha$ such that for all $\beta+1 < \alpha$ we have $X_{\beta} \reduces^{r}_1 X_{\beta+1}$. We will prove that if $\State \setminus \SN \not = \emptyset$, then there is some limit reduction sequence of length $\omega_1$ over $\State \setminus \SN$. Then we will prove that limit reduction sequence of length $\omega_1$ over $\State$ (and with more reason, over $\State \setminus \SN$) cannot exists. The conclusion will be that $\State \setminus \SN = \emptyset$, as wished.

If $X \in \State \setminus \SN$, then there is some infinite reduction sequence $$X = X_0 \reduces^{r}_1 X_1 \reduces^{r}_1 \ldots \reduces^{r}_1 X_n \ldots $$ from $X$. Thus, there is some $X_1$ such that $X \reduces^{r}_1 X_1$ and there is some infinite reduction sequence from $X_1$, hence $X \reduces^{r}_1 X_1$ for some $X_1 \in \State \setminus \SN$. By choice axiom, there is some choice map $$\next: (\State \setminus \SN) \rightarrow (\State \setminus \SN) $$ such that $X \reduces^{r}_1 \next(X)$ for all $X \in \State$. $\next$ is the empty map if $\SN = \State$. From now on, we assume to be fixed a choice map $\next$ as above.

Using $\next$, from any $X \in \State \setminus \SN$ we may easily define an infinite reduction sequence $\next^n(X)$ all in $\State \setminus \SN$. We will prove that we may extend it to a limit reduction sequence on $\State \setminus \SN$ of length $\omega_1$. This is because closed sets in the State Topology are closed by limit, and $\State \setminus \SN$ is a closed set.

In this part of the proof we need the notion of ``definitively true''.

\begin{definition}[Definitively true]
\label{definition-definitively}
Assume $\lambda \in \ON$ is limit and $\sigma = \{X_{\beta} | \beta < \lambda\}$ is any sequence of length $\lambda$.
\begin{enumerate}
\item
$\sigma$ satisfies $X_{\gamma} \subseteq X_{\gamma+1}$ definitively if $\exists \beta<\alpha.\forall \gamma \in [\beta, \lambda[.X_{\gamma} \subseteq X_{\gamma+1}$.
\item
$\sigma$ is definitively weakly increasing if $\exists \beta<\alpha.\forall \gamma, \delta \in [\beta, \lambda[.(\gamma \le \delta) \implies X_{\gamma} \subseteq X_{\delta}$.
\item
$\sigma$ is definitively constant if $\exists \beta<\alpha.\forall \gamma \in [\beta, \lambda[.X_{\beta} = X_{\gamma}$.
\end{enumerate}
\end{definition}

The next step is to prove some easy properties of limit reduction sequences.

\begin{lemma}[Limit Reduction sequences]
\label{lemma-sequences}
Assume $\lambda \in \ON$ is a limit ordinal and $\sigma = \{X_{\alpha} | \alpha < \lambda\}$ is any limit sequence on $\State$ of length $\lambda$. Let $L = \lim_{\gamma \rightarrow \omega_1} X_{\gamma}$.
\begin{enumerate}
\item
If for some $\alpha < \lambda$ and all $\alpha \le \beta < \lambda$ we have $X_\alpha \subseteq X_{\beta}$, then $X_{\alpha} \subseteq L$.

\item
If for some $\alpha < \lambda$ and all $\alpha \le \beta < \lambda$ we have $X_\beta \subseteq X_{\beta+1}$, then $\sigma$ is weakly increasing from the same $\alpha$.

\item
If $\sigma$ is definitively increasing and $\lambda = \omega_1$, then $\sigma$ is definitively stationary.

\item
For any $n \in \Nat$, $\sigma\restr_{<n} = \{X_{\alpha}\restr_{<n} | \alpha < \lambda\}$ is a limit sequence.
\end{enumerate}
\end{lemma}
{\bf Proof}
\begin{enumerate}
\item
Assume $X_{\alpha} \subseteq X_{\gamma}$ for all $\alpha \le \gamma < \lambda$. Then $X_{\alpha} \subseteq \bigcap_{\alpha \le \gamma < \lambda} X_{\gamma} \subseteq \lim X_{\gamma \rightarrow \lambda} X_{\gamma} = L$.

\item
Assume $\alpha \le \alpha' < \lambda$. We prove $X_{\alpha'} \subseteq X_{\beta}$ by induction on $\alpha' \le \beta < \lambda$. Assume $\beta = \alpha'$. Then $X_{\alpha'} \subseteq X_{\alpha'}$. Assume $\beta = \gamma+1 > \gamma \ge \alpha' \ge \alpha$. Then $X_{\alpha'} \subseteq X_{\gamma}$ by induction hypothesis and $X_{\gamma} \subseteq X_{\gamma+1}$ by hypothesis, hence $X_{\alpha'} \subseteq X_{\beta}$. Assume $\beta$ is limit: then $X_{\alpha'} \subseteq X_{\gamma}$ for all $\alpha' \le \gamma < \beta$ by induction hypothesis, therefore $X_{\alpha'} \subseteq X_{\beta}$ by point $1$ applied to the sequence $\{X_{\gamma} | \gamma < \beta\}$.

\item
Assume that $\sigma$ is definitively increasing from some $\alpha$ and $\lambda = \omega_1$, in order to prove that $\sigma$ is definitively stationary. For all $a \in L$ we have $a \in X_{\gamma}$ definitively, therefore there is a first $\xi_a < \omega_1$ such that $a \in X_{\xi_a}  \subseteq X_{\gamma}$ for all $\gamma \ge \xi_a$. Let $\xi$ be l.u.b. of $\{\xi_a | a \in L\} \cup \{\alpha\}$. $L$ is at most countable because $L \subseteq \Atom$, which is at most countable, and $\alpha$ and all $\xi_a$ are $ < \omega_1$, therefore $\xi < \omega_1$. We proved that there is some $\alpha \le \xi < \omega_1$ such that for all $\alpha \le \xi \le \gamma < \omega_1$ we have $L \subseteq X_{\gamma}$. From point $1$ and $X_{\gamma} \subseteq X_{\delta}$ for all $\gamma \le \delta < \omega_1$ we have $X_{\gamma} \subseteq L$. We conclude $L = X_{\gamma}$ for all $\xi \le \gamma < \omega_1$.

\item
Assume $\mu < \lambda$ is limit. Then $X_{\mu} = \bigcup_{\alpha < \mu} \bigcap_{\alpha \le \beta < \mu} X_{\beta}$, hence $X_{\mu} \restr_{<n} = \bigcup_{\alpha < \mu} \bigcap_{\alpha \le \beta < \mu} X_{\beta} \restr_{<n}$. Thus, $\sigma \restr_{<n}$ is a limit sequence.

\end{enumerate}

We explain now how to define a length $\omega_1$ limit reduction sequence in $\State \setminus \SN$. The crucial remark is the following: for any answer in any element of a limit reduction sequence, either the answer belongs to the limit of the sequence together with all answers of level less or equal, or in some future step the is erased together with all answers of the same level (see the first point of the next Lemma).

\begin{lemma}
Assume $\lambda \in \ON$ is a limit ordinal and $\sigma = \{X_{\beta} | \beta < \lambda\}$ is any limit reduction sequence of length $\lambda$. Let $L = \lim_{\beta \rightarrow \lambda} X_\beta \in \State$, and $n \in \Nat$
\begin{enumerate}
\item
For all $\alpha < \lambda$ and all $n \in \Nat$, either $X_{\alpha} \restr_{<n+1} \subseteq L$, or there is some $\alpha < \gamma < \lambda$ such that $X_{\gamma}\restr_{=m} = \emptyset$
\item
$L$ is topologically adherent to $\{X_{\beta} | \beta < \lambda\}$ (that is, any open set including $L$ intersects $\{X_{\beta} | \beta < \lambda\}$.
\item
If $C \subseteq \State$ is closed and $\{X_{\beta} | \beta < \lambda\} \subseteq C$ then $L \in C$
\item
$\State \setminus \SN$ is closed
\item
If $\State \setminus \SN \not = \emptyset$, then there is some length $\omega_1$ limit reduction sequence in $\State \setminus \SN$.
\end{enumerate}
\end{lemma}

{\bf Proof}
\begin{enumerate}
\item
Consider the sequence $\tau = \{X_{\beta} \restr_{<n+1} | \beta < \lambda\}$: this is a limit sequence by Lemma \ref{lemma-sequences}.3. If $X_{\beta} \restr_{<n+1} \subseteq X_{\beta+1} \restr_{<n+1}$ for all $\alpha \le \beta < \lambda$, then $\tau$ is weakly increasing from $\alpha$ by Lemma \ref{lemma-sequences}.1. In this case $X_{\alpha} \restr_{<n+1} \subseteq X_{\gamma} \restr_{<n+1} \subseteq X_{\gamma} $ for all $\alpha \le \gamma < \lambda$, therefore $X_{\alpha} \restr_{<n+1} \subseteq L$ by definition of $L$. Assume instead that $X_{\beta} \restr_{<n+1} \not \subseteq X_{\beta+1} \restr_{<n+1}$ for some $\alpha \le \beta < \lambda$. Then by Lemma \ref{lemma-reduction}.4 we have $X_{\beta+1} \restr_{=n} = \emptyset$.

\item
Fix $\alpha < \lambda$, and assume $O$ is any sub-basic open and $L \in O$, in order to prove that $X_{\beta} \in O$ for some $\alpha \le \beta < \lambda$. For some $a \in \Atom$, either $O = A_a = \{X \in \State | a \in X\}$, or $O = B_a = \{X \in \State | X \cap [a]_{\sim} = \emptyset \}$. We reason by cases.
\begin{enumerate}
\item
If $O = A_a$ we have $a \in L$. By definition of $L$, for some $\alpha < \lambda$ and all $\alpha \le \beta < \lambda$ we have $a \in X_{\beta}$. In particular, $a \in X_{\alpha}$, hence $X_{\alpha} \in O$.
\item
If $O = B_a$ we have  $L \cap [a]_{\sim} = \emptyset$. Assume $n = \Level(a)$: by point $1$ above there is some $\alpha \le \beta < \lambda$ such that either $X_{\beta} \restr_{<n+1} \subseteq L$ or $X_{\beta} \restr_{=n} = \emptyset$. In both cases we have $X_{\beta} \restr_{=n} \subseteq L \restr_{=n}$, either because $X_{\beta} \restr_{=n} = (X_{\beta} \restr_{<n+1}) \restr_{=n} \subseteq L\restr_{=n}$, or because $X_{\beta} \restr_{=n} = \emptyset \subseteq L\restr_{=n}$. We deduce $X_{\beta} \cap [a]_{\sim} = (X_{\beta} \restr_{=n}) \cap [a]_{\sim} \subseteq (L \restr_{=n}) \cap [a]_{\sim} \subseteq L \cap [a]_{\sim} = \emptyset$. Thus, $X_{\beta} \in B_a$.
\end{enumerate}

\item
Assume $C \subseteq \State$ is closed and $\{X_{\beta} | \beta < \lambda\} \subseteq C$ in order to prove that $L \in C$. Assume for contradiction that $L \not \in C$. Then $L \in \State \setminus C$, which is open. By the previous point we have $X_{\alpha} \in \State \setminus C$ for some $\alpha < \lambda$, contradicting $X_{\alpha} \in C$.

\item
$\State \setminus \SN$ is closed because $\SN$ is open.

\item
From any $X \in \State \setminus \SN$ we may define a limit reduction sequence of length $\omega_1$ (and in fact of {\em any} length). We set $X_0 = X$, $X_{\alpha+1} = \next(X_{\alpha})$ for all $\alpha < \omega_1$ and $X_{\lambda} = \lim_{\beta \rightarrow \lambda} X_{\beta}$ for all limit $\lambda < \omega_1$. We check that the definition is correct. By assumption $X_0 = X \in \State \setminus \SN$. Assume $\alpha < \omega_1$ and $X_{\alpha} \in \State \setminus \SN$, then $X_{\alpha+1} = \next(X_{\alpha}) \in \State \setminus \SN$. Assume $\lambda < \omega_1$ is limit and $\{X_{\beta} | \beta < \lambda\} \subseteq \State \setminus \SN$. Since $\State \setminus \SN$ is closed by point $3$, then by point $2$ above we have $X_{\lambda} = \lim_{\beta \rightarrow \lambda} X_{\beta} \in \State \setminus \SN$.
\end{enumerate}

\section{A termination result from an algorithm searching fixed points}
\label{section-termination}
In the previous section we proved that if $\State \setminus \SN \not = \emptyset$, then there is a limit reduction sequence of length $\omega$. In this section we will prove that no limit reduction sequence of length $\omega_1$ may exists, and we will conclude that $\State \setminus \SN = \emptyset$, as wished. We first prove that limit reduction sequences of length $\omega_1$ are definitively stationary.

\begin{lemma}[Stationarity]
\label{lemma-limit}
Assume $\sigma = \{X_{\alpha} | \alpha < \omega_1\}$ is any sequence on $\State$. Let $n \in \Nat$.
\begin{enumerate}
\item
For any limit reduction sequence $\{X_{\beta} | \beta < \omega_1\}$ of length $\omega_1$ on $\State$, the sequence $\{X_{\beta} \restr_{<n}| \beta < \omega_1\}$ is definitively stationary.
\item
Any limit reduction sequence $\{X_{\beta} | \beta < \omega_1\}$ of length $\omega_1$ on $\State$ is definitively stationary.
\end{enumerate}
\end{lemma}

{\bf Proof}
\begin{enumerate}
\item
We argue by induction on $n \in \Nat$. Assume $n=0$: then $X_{\beta} \restr_{<n} = \emptyset$ is definitively stationary. Assume $X_{\beta} \restr_{<n}$ is definitively stationary, in order to prove $X_{\beta} \restr_{<n+1}$ is definitively stationary. If $X_{\beta} \restr_{<n} = X_{\beta+1} \restr_{<n}$ then $X_{\beta} \restr_{<n+1} \subseteq X_{\beta+1} \restr_{<n}$ by Lemma \ref{lemma-reduction}.6, hence we definitively have $X_{\beta} \restr_{<n+1} \subseteq X_{\beta+1} \restr_{<n}$. By Lemma \ref{lemma-sequences}.4 $X_{\beta} \restr_{<n+1}$ is a limit sequence, and by Lemma \ref{lemma-sequences}.2 it is weakly increasing. It has length $\omega_1$, therefore by Lemma \ref{lemma-sequences}.3 it is definitively stationary.

\item
By the previous point, for all $n \in \Nat$ there is a first $\alpha_n < \omega_1$ such that $X_{\beta} \restr_{<n}$ is stationary from $\alpha_n$. Let $\alpha < \omega_1$ by the l.u.b. of $\{\alpha_n | n \in \Nat\}$: then for all $n \in \Nat$, $X_{\beta} \restr_{<n}$ is stationary from $\alpha$. Thus, $X_{\beta} $ is stationary from $\alpha$.
\end{enumerate}

The strong termination result for the reduction relation $\reduces^{r}_1$ easily follows.

\begin{theorem}[Pre-fixed point Theorem]
For all states $X \in \State$, for all realizers $r : \State \rightarrow \FinPow(\Atom)$, all reduction sequences $X \reduces^{r}_1 X_1 \reduces^{r}_1 \ldots$ $\ldots \reduces^{r}_1 X_n \reduces^{r}_1 \ldots$ from $X$ are finite.
\end{theorem}

{\bf Proof}
Assume there is some $X \in \State \setminus \SN$. By Lemma \ref{lemma-limit}.5 there is some limit reduction sequence $\{X_{\beta} | \beta < \omega_1\} \subseteq (\State \setminus \SN)$ from $X$ of length $\omega_1$. By Lemma \ref{lemma-limit}.2, $\{X_{\beta} | \beta < \omega_1\}$ is definitively stationary, therefore for some $\alpha < \omega_1$ we have $X_{\alpha+1} = X_{\alpha}$, hence $X_{\alpha} \reduces^{r}_1 X_{\alpha+1} = X_{\alpha}$, against Lemma \ref{lemma-reduction}.2.

\section{Related works and conclusions}\label{relatedWorks}

In this section we stress the most relevant differences of the present work w.r.t. the ones by the authors themselves and by others. The essential difference w.r.t. \cite{AB} and \cite{BerardiLiguoroMonadi} is non-monotonicity.
In\cite{AschieriPhD} also the case of non-monotonic learning is considered, though only deterministic learning processes are treated.
In \cite{LMCS2013}, which is the full version of \cite{full-abstract}, we propose a deterministic algorithm to compute a (finite) sound pre-fixed point of any effective realizer; however we have been able to treat the case in which the maximum level of answers is $2$, while here we have a termination proof of a non-deterministic algorithm working on states with answers of arbitrary level $< \omega$.

We stress that non-determinism is no minor trick. First, if the output $r(X)$ of a realizer may include more than one answer, then our convergence result also holds for any $r' : \State \rightarrow \FinPow(\Atom)$, {\em even if not continuous}, provided there is some continuous $r : \State \rightarrow \FinPow(\Atom)$ such that $r'(X) \subseteq r(X)$ for all $X \in \State$. This simple remark shows that the result for the non-deterministic case is much stronger than the result for the deterministic one.

In \cite{Mints} Mints considered the $\omega$-level version of the problem. In our terminology, he introduced a non-deterministic reduction relation adding one answer at the time, and proved a {\em weak} normalization result: there is a reduction sequence from the empty state to some sound irreducible state. However, in \cite{Mints} there is no normalizing reduction strategy, and we suspect that the strong normalization result would fail in that setting.

In conclusion we have presented a new result that we consider as a step toward a realistic use of non constructive proofs as algorithms. Improvements are certainly possible, such as for example a more sophisticated way of representing logical dependencies than level. The aim is to find an algorithm removing the minimum amount of answers from a state when adding new ones, hence resulting into a faster computation.

\pagestyle{empty}

\nocite{*}
\bibliographystyle{eptcs}
\bibliography{generic}
\end{document}